\documentclass[conference]{IEEEtran}

\ifCLASSINFOpdf
\else
\fi
\usepackage{url}

\usepackage[]{fancyhdr} %
\newcommand{\changefont}{\fontsize{9}{9}\selectfont}
\usepackage{bm}
\usepackage{cite}
\usepackage{amsmath,amssymb,amsfonts}
\usepackage{algorithmic}
\usepackage{array}
\usepackage{multirow}
\usepackage{threeparttable}
\usepackage{graphicx}
\usepackage{textcomp}
\usepackage{xcolor}
\usepackage{makecell}
\IEEEoverridecommandlockouts
\begin{document}

%
\title{ATC-Based Scenario Decomposition Algorithm for Optimal Power Flow of Distribution Networks Considering High Photovoltaic Penetration\\
\thanks{This work is supported by the National Natural Science Foundation of China through Project No. 62173287.}
}

\author{\IEEEauthorblockN{Xiemin Mo, Tao Liu}
\IEEEauthorblockA{1. Department of Electrical and Electronic Engineering\\
The University of Hong Kong\\
Hong Kong SAR, China \\
2. Shenzhen Institute of Research and Innovation\\
The University of Hong Kong\\
\{xieminmo, taoliu\}@eee.hku.hk}
\and
\IEEEauthorblockN{Xue Lyu}
\IEEEauthorblockA{Department of Electrical and Computer Engineering \\
The University of Wisconsin-Madison\\
Madison, WI, U.S.A \\
xuelu111@gmail.com}
}


%





\maketitle
\thispagestyle{fancy}
\pagestyle{fancy}


\begin{abstract}
  This paper focuses on the analytical target cascading (ATC) based scenario decomposition method which applies to the stochastic OPF problem of distribution networks with high photovoltaic penetration. The original two-stage stochastic OPF model is decomposed into a master problem in the upper level and multiple subproblems in the lower level. This decomposition makes subproblems easier to be solved and can also effectively overcome the curse of dimensionality in the traditional scenario-based model. The global optimal solution can be obtained by only transferring some necessary coupling information between the upper and lower levels. Moreover, all the subproblems in the lower level can be solved in a parallel manner which improves the computational efficiency, in particular, for cases with a larger number of scenarios. Case studies on the IEEE 33-bus system and various larger systems verify the effectiveness and adaptability of the proposed algorithm.
  \end{abstract}

\begin{IEEEkeywords}
    Optimal power flow, stochastic programming, analytical target cascading
\end{IEEEkeywords}


%
\IEEEpeerreviewmaketitle

\section{Introduction}
\IEEEPARstart{A}{s} an important form for the integration of distributed renewable energy, distribution networks have attracted increasing attention in the last decade. Due to the strong uncertainty of solar energy, the economic operation of a distribution network will face more severe challenges as the installment of photovoltaic panels (PVs) increasing \cite{Alyami2014}. Therefore, it is necessary to study the stochastic optimal power flow (OPF) problem of distribution networks.\\
\indent{As a popular method in the field of stochastic programming, the scenario-based method has been widely used in power systems. Firstly, a curve is selected as the forecasted scenario of random variables according to the system operation experience. Based on this scenario, a set of random error scenarios is generated by utilizing the sampling method. Then the model based on the combination of the forecasted scenario and the error scenario set is solved to obtain the expected optimal solution which is capable of fully considering the uncertainties in the future \cite{Wu2007}. However, a large number of error scenarios will inevitably cause the curse of dimensionality in the application. Therefore, it is crucial to reduce the model complexity with the satisfaction of decision accuracy.}\\
\indent{Firstly, it is an intuitive way to select and reduce error scenarios of random variables. In \cite{Fu2020}, based on the nearest neighbor theory and nonnegative matrix decomposition, a reduction method is developed to effectively cut down the dimension of random variables and the size of error scenarios set. In \cite{Li2016}, a procedure of formulating strategies for reducing error scenarios is creatively formulated as a mathematical optimization problem containing stochastic variables in time series. In order to further consider the correlation between random variables, a scenario tree model is adopted to describe the transfer relationship between different random variables \cite{Beltran2017}. Then, the size of the scenario tree is decreased by using the reduction algorithm with quadratic process distance. Different from deleting scenarios directly, a clustering method which aims to combine several scenarios together is proposed for scenario reduction based on the concept of correlation loss in \cite{Hu2019}.}\\
\indent{Another possible way to reduce the model size is to embed the scenario method into the framework of the spatial decomposition-based distributed algorithm. In \cite{Ahmadi2013}, the optimality condition of the centralized problem is decomposed into several small subproblems corresponding to dispatch areas. Then, the scenario method is used to construct the stochastic programming model in each area. In \cite{Ahmadi2014}, the decomposition of the centralized problem is conducted by using auxiliary problem principle and the stochastic programming model of each area is constructed in a similar way as that in \cite{Ahmadi2013}.}\\
\indent{However, the scenario-based model is a centralized problem with multiple scenarios in the above-mentioned works. Hence, the problem of the dimensionality curse caused by a large number of scenarios remains open. This paper focuses on the analytical target cascading (ATC) based scenario decomposition method which applies to the stochastic OPF problem of distribution networks with high photovoltaic penetration. Compared with the existing works, the proposed method has the following contributions:}\\
\indent{1) This paper studies the method of reducing model complexity from the aspect of scenario decomposition. Based on ATC, the proposed method decomposes the original scenario-based model into a series of small-scale models of two levels, which dramatically reduces the complexity of the model and can effectively overcome the curse of dimensionality.}\\
\indent{2) Different from the existing methods, all the stochastic scenarios can be fully considered by transferring moderate necessary coupling information between the upper-level and lower-level problems. Moreover, a proposed parallel computing manner of the lower-level problems can obviously improve the computational efficiency with the increasing number of stochastic scenarios, which also demonstrates the good application potentiality.}\\
\indent{The rest of this paper is organized as follows. Section II presents the problem formulation of the studied stochastic OPF model. Section III describes the proposed ATC-based scenario decomposition method. Case studies on the IEEE 33-bus system and several larger systems are shown in Section IV. Finally, Section V presents the conclusions.}\\
\vspace{-0.3cm}
\section{Problem Formulation}
\indent{In this section, the scenario-based two-stage stochastic OPF model 
is proposed to minimize the expected operation cost by finding the optimal dispatch decisions under the uncertainties of PVs and loads within time period $t\in \mathcal{T}$.
In this model, the first stage corresponds to the forecasted scenario and the second stage corresponds to the error scenarios.}\\
\indent{For brevity, the proposed model can be represented in the following compact form:}\\
\begin{subequations}
   \begin{equation}
      \min C^{f}(\bm{x}_{t}^{f})+\sum_{s \in \mathcal{S}}\pi_{s}C^{sec}_{s}(\bm{x}_{s,t}^{sec}) \label{eq:eq1a}
   \end{equation}
   \begin{equation}
      s.t.\quad
      \bm{g}^{f}(\bm{x}_{t}^{f}) \preceq \bm{0}, \forall t \in \mathcal{T} \label{eq:eq1b}
   \end{equation}
   \begin{equation}
      \bm{g}^{sec}_{s}(\bm{x}_{s,t}^{sec}) \preceq \bm{0}, \forall t \in \mathcal{T}, \forall s \in \mathcal{S} \label{eq:eq1c}
   \end{equation}
   \begin{equation}
      \bm{g}^{co}_{s}(\bm{x}_{t}^{f},\bm{x}_{s,t}^{sec}) \preceq \bm{0}, \forall t \in \mathcal{T}, \forall s \in \mathcal{S} \label{eq:eq1d}
   \end{equation}
\end{subequations}
\noindent{where $\bm{x}_{t}^{f},\bm{x}_{s,t}^{sec}$ are the decision vectors of the first and second stage, respectively; $C^{f}(\cdot), C^{sec}_{s}(\cdot)$ in (1a) are the cost functions of the first stage under forecasted scenario and second stage under error scenarios $s \in \mathcal{S}$, respectively; $\pi_{s}$ is the probability of error scenario $s$; $\bm{g}^{f}(\cdot),\bm{g}^{sec}_{s}(\cdot)$ represents the functions of the first and second stage variables, respectively; $\bm{g}^{co}_{s}(\cdot)$ represents the coupling function between the first and second stage variables.}\\
\indent{Equation (\ref{eq:eq1a}) represents the total cost. Equation (\ref{eq:eq1b}) represents the first stage constraints. Equation (\ref{eq:eq1c}) represents the second stage constraints. Equation (\ref{eq:eq1d}) represents the coupling constraints between the first and second stages.}\\
\indent{Consider a radial distribution network with $|\mathcal{N}|$ buses and $|\mathcal{L}|$ lines. Denote that bus index $i,j,k \in \mathcal{N}$ and line index $ij \in \mathcal{L}$. Assume that each bus $i$ can be equipped with one or more devices, such as distributed generators (DGs), PVs and loads. $\mathcal{G}$ denotes the set of all generators (i.e., DGs and PVs). Then, the corresponding detailed formulations are given as below:}\\
\indent{\textit{1) Objective Function}: Equation (\ref{eq:eq1a}) aims to minimize the total cost of the first and second stages. In the first stage, the cost includes power losses and the PV curtailment penalty cost under the forecast scenario. 
In the second stage, the cost includes power losses, the PV curtailment penalty cost, and re-dispatch cost of DGs under the error scenarios.}\\
\begin{equation}
   \begin{aligned}
   &\min \sum_{t \in \mathcal{T}}\left(\right. \sum_{ij \in \mathcal{L}}c_{loss}r_{ij}l_{ij,t}^{f}+\sum_{i \in \mathcal{G}}c_{pv}(P^{av,f}_{PV,i,t}-P_{PV,i,t}^{f})\left.\right) \\
   &+\sum_{s \in \mathcal{S}}\pi_{s}\left\{\right.\sum_{t \in \mathcal{T}}\left(\right.  \sum_{ij \in \mathcal{L}}c_{loss}r_{ij}l_{s,ij,t}^{sec}+\sum_{i\in \mathcal{G}}c_{pv}(P^{av,sec}_{PV,s,i,t}\\
   &-P_{PV,s,i,t}^{sec})+\sum_{i \in \mathcal{G}}c_{DG,i}^{sec}(P_{DG,s,i,t}^{+,sec}+P_{DG,s,i,t}^{-,sec}+Q_{DG,s,i,t}^{+,sec}\\
   &+Q_{DG,s,i,t}^{-,sec})\left.\right)\left.\right\} \label{eq:eq2}
   \end{aligned}
\end{equation}
\noindent{where $c_{loss},c_{pv},$ and $c_{DG,i}^{sec}$ are linear cost coefficients of power losses, PV curtailment penalty, and DG re-dispatch, respectively;
         $r_{ij}$ is the resistance of line $ij$; $l_{ij,t}^{f},l_{s,ij,t}^{sec}$ are the square of current of the first and second stages, respectively;
         $P_{PV,i,t}^{f},P_{PV,s,i,t}^{sec}$ are active power outputs of PVs for the first and second stages, respectively;
         $P_{DG,s,i,t}^{+,sec},P_{DG,s,i,t}^{-,sec},Q_{DG,s,i,t}^{+,sec},Q_{DG,s,i,t}^{-,sec}$ are active and reactive re-dispatch power outputs at the second stage, respectively;
         $P_{PV,i,t}^{av,f}, P_{PV,s,i,t}^{av,sec}$ are available PV outputs of the first and second stages, respectively;}\\
\indent{\textit{2) Decision Variables of the First Stage}: $\bm{x}_{t}^{f}$ can be described in details as follows:}
\begin{equation}
   \begin{aligned}
   \bm{x}_{t}^{f} = \left.\right\{ &P_{DG,i,t}^{f},Q_{DG,i,t}^{f},P_{PV,i,t}^{f},Q_{PV,i,t}^{f},P_{i,t}^{f},Q_{i,t}^{f},\\
   &T_{P,ij,t}^{f},T_{Q,ij,t}^{f},l_{ij,t}^{f},v_{i,t}^{f}  \left.\right\}, \forall i \in \mathcal{N}, \forall ij \in \mathcal{L}, \forall t \in \mathcal{T}  \label{eq:eq3}
   \end{aligned}
\end{equation}
\noindent{where $P_{DG,j,t}^{f},Q_{DG,j,t}^{f}$ are active and reactive power outputs of DGs of the first stage, respectively; $Q_{PV,i,t}^{f}$ is the reactive power output of PV for the first stage; $P_{i,t}^{f},Q_{i,t}^{f}$ are nodal active and reactive power injections of the first stage, respectively; $T_{P,ij,t}^{f},T_{Q,ij,t}^{f}$ are active and reactive power flows of the first stage, respectively; $v_{i,t}^{f}$ is the square of voltage of the first stage.}\\
\indent{\textit{3) Constraints of the First Stage}: In the forecasted scenario, equation (\ref{eq:eq1b}) includes the nodal active and reactive power balance constraints (\ref{eq:eq4})-(\ref{eq:eq7}), the voltage drop constraints (\ref{eq:eq8}), the voltage security limits (\ref{eq:eq9}), the system reference voltage constraint (\ref{eq:eq10}), the active and reactive power flow limits (\ref{eq:eq11})-(\ref{eq:eq12}), the active and reactive power outputs limits of DGs (\ref{eq:eq13})-(\ref{eq:eq14}), the active and reactive power outputs limits of PVs (\ref{eq:eq15})-(\ref{eq:eq16}), and the second order cone programming (SOCP) constraints for line $ij$ (\ref{eq:eq17}). Note that the above SOCP relaxation is equivalent to the original nonlinear constraints before relaxation for radial distribution network \cite{Farivar2013}, \cite{Zheng2016}. The above-mentioned constraints for $\forall t \in \mathcal{T}$ are given in details as below: }\\
\begin{equation}
   \begin{aligned}
   P_{j,t}^{f} + \sum_{i:i \rightarrow j}(T_{P,ij,t}^{f}-l_{ij,t}^{f}r_{ij}) = \sum_{j:j \rightarrow k}T_{P,jk,t}^{f}, \forall j \in \mathcal{N} \label{eq:eq4}   
   \end{aligned}
\end{equation}
\begin{equation}
   \begin{aligned}
  Q_{j,t}^{f} + \sum_{i:i \rightarrow j}(T_{Q,ij,t}^{f}-l_{ij,t}^{f}x_{ij}) = \sum_{j:j \rightarrow k}T_{Q,jk,t}^{f}, \forall j \in \mathcal{N} \label{eq:eq5}   
   \end{aligned}
\end{equation}
\begin{equation}
   P_{j,t}^{f}=\left \{
\begin{aligned} 
   &P_{DG,j,t}^{f}+P_{PV,j,t}^{f}-P_{L,j,t}^{f}, \forall j \in \mathcal{G} \\
   &-P_{L,j,t}^{f}, \forall j \in \mathcal{N} \label{eq:eq6}
\end{aligned} \right.
\end{equation}
\begin{equation}
   \hspace{0.15cm}Q_{j,t}^{f}=\left \{
\begin{aligned} 
   &Q_{DG,j,t}^{f}+Q_{PV,j,t}^{f}-Q_{L,j,t}^{f}, \forall j \in \mathcal{G} \\
   &-Q_{L,j,t}^{f}, \forall j \in \mathcal{N} \label{eq:eq7}
\end{aligned} \right.
\end{equation}
\begin{equation}
   \begin{aligned}
   v_{j,t}^{f} = v_{i,t}^{f} - 2(r_{ij}T_{P,ij,t}^{f}+x_{ij}T_{Q,ij,t}^{f}) + (r_{ij}^{2}+x_{ij}^{2})l_{ij,t}^{f}, \\
   \forall ij \in \mathcal{L} \label{eq:eq8}   
   \end{aligned}
\end{equation}
\begin{equation}
   \underline{V}_{i}^{2} \le v_{i,t}^{f} \le \overline{V}_{i}^{2}, \forall i \in \mathcal{N} \label{eq:eq9}
\end{equation}
\begin{equation}
   v_{1,t}^{f} = V_{ref}^{2} \label{eq:eq10}
\end{equation}
\begin{equation}
   \underline T_{P,ij} \le T_{P,ij,t}^{f} \le \overline T_{P,ij},\forall ij \in \mathcal{L} \label{eq:eq11}
\end{equation}
\begin{equation}
   \underline T_{Q,ij} \le T_{Q,ij,t}^{f} \le \overline T_{Q,ij},\forall ij \in \mathcal{L} \label{eq:eq12}
\end{equation}
\begin{equation}
   \underline P_{DG,i} \le P_{DG,i,t}^{f} \le \overline P_{DG,i},\forall i \in \mathcal{G} \label{eq:eq13}
\end{equation}
\begin{equation}
   \underline Q_{DG,i} \le Q_{DG,i,t}^{f} \le \overline Q_{DG,i},\forall i \in \mathcal{G} \label{eq:eq14}
\end{equation}
\begin{equation}
   0 \le P_{PV,i,t}^{f} \le P^{av,f}_{PV,i,t},\forall i \in \mathcal{G} \label{eq:eq15}
\end{equation}
\begin{equation}
   \lvert Q_{PV,i,t}^{f} \rvert \le P_{PV,i,t}^{f}\tan\phi,\forall i \in \mathcal{G} \label{eq:eq16}
\end{equation}
\begin{equation}
 ||2T_{P,ij,t}^{f},2T_{Q,ij,t}^{f},l_{ij,t}^{f}-v_{i,t}^{f}||_{2}\le l_{ij,t}^{f}+v_{i,t}^{f},\forall ij \in \mathcal{L} \label{eq:eq17}
\end{equation}
\noindent{where $\overline P_{DG,i},\underline P_{DG,i}$ are maximum and minimum active power outputs, respectively; $\overline Q_{DG,i},\underline Q_{DG,i}$ are maximum and minimum reactive power outputs, respectively; $\overline T_{P,ij}, \underline T_{P,ij}$ are active transmission limits; $\overline T_{Q,ij}, \underline T_{Q,ij}$ are reactive transmission limits; $x_{ij}$ is the reactance of line $ij$; $\overline{V}_{i}^{2},\underline{V}_{i}^{2}$ are maximum and minimum of the square of voltage magnitudes, respectively; $V_{ref}$ is the voltage of the reference bus; $\phi$ is the power
factor angle; $P_{L,j,t}^{f},Q_{L,j,t}^{f}$ are active and reactive loads of the first and second stages, respectively;}\\
\indent{\textit{4) Decision Variables of the Second Stage}: $\bm{x}_{s,t}^{sec}$ can be characterized as follows:}
\begin{equation}
   \begin{aligned}
   \bm{x}_{s,t}^{sec} = \left.\right\{ &P_{DG,s,i,t}^{sec},Q_{DG,s,i,t}^{sec},P_{DG,s,i,t},Q_{DG,s,i,t},\\
   &P_{DG,s,j,t}^{+,sec},Q_{DG,s,j,t}^{+,sec},P_{DG,s,j,t}^{-,sec},Q_{DG,s,j,t}^{-,sec},\\
   &P_{PV,s,i,t}^{sec},Q_{PV,s,i,t}^{sec},P_{s,j,t}^{sec},Q_{s,j,t}^{sec},T_{P,s,ij,t}^{sec},T_{Q,s,ij,t}^{sec},\\
   &l_{s,ij,t}^{sec}, v_{s,i,t}^{sec}  \left.\right\},\forall i \in \mathcal{N}, \forall ij \in \mathcal{L}, \forall t \in \mathcal{T}  \label{eq:eq18}
   \end{aligned}
\end{equation}
\noindent{where $P_{DG,s,i,t}^{sec},Q_{DG,s,i,t}^{sec}$ are active and reactive power outputs of DGs of the second stages, respectively; $P_{DG,s,i,t},Q_{DG,s,i,t}$ are auxiliary variables to describe the first-stage active and reactive power outputs are identical in each scenario of the second stage; $Q_{PV,s,i,t}^{sec}$ is the reactive power output of PV of the second stage; $P_{s,j,t}^{sec},Q_{s,j,t}^{sec}$ are nodal active and reactive power injections of the second stage, respectively; $T_{P,s,ij,t}^{sec},T_{Q,s,ij,t}^{sec}$ are active and reactive power flows of the second stage, respectively; $v_{s,i,t}^{sec}$ is the square of voltage magnitude of the second stage.}\\
\indent{\textit{5) Constraints of the Second Stage}: In each error scenario $s$ of the second stage, equation (\ref{eq:eq1c}) includes the power outputs of the second stage considering re-dispatch decisions (\ref{eq:eq19})-(\ref{eq:eq20}), the nodal active and reactive power balance constraints (\ref{eq:eq21})-(\ref{eq:eq24}), the voltage drop constraints (\ref{eq:eq25}), the voltage security limits (\ref{eq:eq26}), the system reference voltage constraint (\ref{eq:eq27}), the active and reactive power flow limits (\ref{eq:eq28})-(\ref{eq:eq29}), the active and reactive power outputs limits of DGs (\ref{eq:eq30})-(\ref{eq:eq31}), the active and reactive power outputs limits of PVs (\ref{eq:eq32})-(\ref{eq:eq33}), the SOCP constraints for line ij (\ref{eq:eq34}), the up and down limits of active re-dispatch outputs (\ref{eq:eq35})-(\ref{eq:eq36}), and the up and down limits of reactive re-dispatch power outputs (\ref{eq:eq37})-(\ref{eq:eq38}). The above-mentioned constraints for $\forall t \in \mathcal{T}$ and $\forall s \in \mathcal{S}$ are given in details as below:}
\begin{equation}
   P_{DG,s,j,t}^{sec}=P_{DG,s,j,t}+P_{DG,s,j,t}^{+,sec}-P_{DG,s,j,t}^{-,sec},\forall j \in \mathcal{G} \label{eq:eq19}
\end{equation}
\begin{equation}
   \hspace{0.25cm}Q_{DG,s,j,t}^{sec}=Q_{DG,s,j,t}+Q_{DG,s,j,t}^{+,sec}-Q_{DG,s,j,t}^{-,sec},\forall j \in \mathcal{G} \label{eq:eq20}
\end{equation}
\begin{equation}
   \begin{aligned}
   P_{s,j,t}^{sec} + \sum_{i:i \rightarrow j}(T_{P,s,ij,t}^{sec}-l_{s,ij,t}^{sec}r_{ij}) = \sum_{j:j \rightarrow k}T_{P,s,jk,t}^{sec}, \forall j \in \mathcal{N} \label{eq:eq21}   
   \end{aligned}
\end{equation}
\begin{equation}
   \begin{aligned}
  Q_{s,j,t}^{sec} + \sum_{i:i \rightarrow j}(T_{Q,s,ij,t}^{sec}-l_{s,ij,t}^{sec}x_{ij}) = \sum_{j:j \rightarrow k}T_{Q,s,jk,t}^{sec}, \forall j \in \mathcal{N} \label{eq:eq22}   
   \end{aligned}
\end{equation}
\begin{equation}
   P_{s,j,t}^{sec}=\left \{
\begin{aligned} 
   &P_{DG,s,j,t}^{sec}+P_{PV,s,j,t}^{sec}-P_{L,s,j,t}^{sec}, \forall j \in \mathcal{G} \\
   &-P_{L,s,j,t}^{sec}, \forall j \in \mathcal{N}\label{eq:eq23}
\end{aligned} \right.
\end{equation}
\begin{equation}
   Q_{s,j,t}^{sec}=\left \{
\begin{aligned} 
   &Q_{DG,s,j,t}^{sec}+Q_{PV,s,j,t}^{sec}-Q_{L,s,j,t}^{sec}, \forall j \in \mathcal{G} \\
   &-Q_{L,s,j,t}^{sec}, \forall j \in \mathcal{N}  \label{eq:eq24}
\end{aligned} \right.
\end{equation}
\begin{equation}
   \begin{aligned}
   v_{s,j,t}^{sec} = v_{s,i,t}^{sec} - 2(r_{ij}T_{P,s,ij,t}^{sec}&+x_{ij}T_{Q,s,ij,t}^{sec})\\
  &+ (r_{ij}^{2}+x_{ij}^{2})l_{s,ij,t}^{sec}, \forall ij \in \mathcal{L} \label{eq:eq25}   
   \end{aligned}
\end{equation}
\begin{equation}
   \underline{V}_{i}^{2} \le v_{s,i,t}^{sec} \le \overline{V}_{i}^{2}, \forall i \in \mathcal{N} \label{eq:eq26}
\end{equation}
\begin{equation}
   v_{s,1,t}^{sec} = V_{ref}^{2} \label{eq:eq27}
\end{equation}
\begin{equation}
   \underline T_{P,ij} \le T_{P,s,ij,t}^{sec} \le \overline T_{P,ij},\forall ij \in \mathcal{L} \label{eq:eq28}
\end{equation}
\begin{equation}
   \underline T_{Q,ij} \le T_{Q,s,ij,t}^{sec} \le \overline T_{Q,ij},\forall ij \in \mathcal{L} \label{eq:eq29}
\end{equation}
\begin{equation}
   \underline P_{DG,i} \le P_{DG,s,i,t}^{sec} \le \overline P_{DG,i},\forall i \in \mathcal{G} \label{eq:eq30}
\end{equation}
\begin{equation}
   \underline Q_{DG,i} \le Q_{DG,s,i,t}^{sec} \le \overline Q_{DG,i},\forall i \in \mathcal{G} \label{eq:eq31}
\end{equation}
\begin{equation}
   0 \le P_{PV,s,i,t}^{sec} \le P^{av,sec}_{PV,s,i,t},\forall i \in \mathcal{G} \label{eq:eq32}
\end{equation}
\begin{equation}
   \lvert Q_{PV,s,i,t}^{sec} \rvert \le P_{PV,s,i,t}^{sec}\tan\phi,\forall i \in \mathcal{G} \label{eq:eq33}
\end{equation}
\begin{equation}
||2T_{P,s,ij,t}^{sec},2T_{Q,s,ij,t}^{sec},l_{s,ij,t}^{sec}-v_{s,i,t}^{sec}||_{2} \le l_{s,ij,t}^{sec}+v_{s,i,t}^{sec},\forall ij \in \mathcal{L} \label{eq:eq34}
\end{equation}
\begin{equation}
   0 \le P_{DG,s,i,t}^{+,sec} \le \overline P_{DG,i}-P_{DG,i,t},\forall i \in \mathcal{G} \label{eq:eq35}
\end{equation}
\begin{equation}
   0 \le P_{DG,s,i,t}^{-,sec} \le P_{DG,i,t},\forall i \in \mathcal{G} \label{eq:eq36}
\end{equation}
\begin{equation}
   0 \le Q_{DG,s,i,t}^{+,sec} \le \overline Q_{DG,i}-Q_{DG,i,t},\forall i \in \mathcal{G} \label{eq:eq37}
\end{equation}
\begin{equation}
   0 \le Q_{DG,s,i,t}^{-,sec} \le Q_{DG,i,t},\forall i \in \mathcal{G} \label{eq:eq38}
\end{equation}
\noindent{where $P_{L,s,j,t}^{sec},Q_{L,s,j,t}^{sec}$ are active and reactive loads of the second stage, respectively;}\\
\indent{\textit{Coupling Constraints}: Equation (\ref{eq:eq1d}) ensures the first-stage power outputs are identical in each scenario of the second stage which includes the following constraints for $\forall i \in \mathcal{G},\forall t \in \mathcal{T},\forall s \in \mathcal{S}$:}\\
\begin{equation}
   P_{DG,i,t}^{f}=P_{DG,s,i,t} \label{eq:eq39}
\end{equation}
\begin{equation}
   Q_{DG,i,t}^{f}=Q_{DG,s,i,t} \label{eq:eq40}
\end{equation}
\indent{To sum up, the optimal solutions which satisfy various constraints (i.e., nodal power balance, power flow limits, and voltage security limits) can be obtained by solving the above-mentioned scenario-based two-stage stochastic OPF model (2)-(40). However, a larger system with more scenarios may make the model computational intractable.}\\
\begin{figure}[!]
	\centering
	\setlength{\abovecaptionskip}{0cm}
	\includegraphics[scale=0.39]{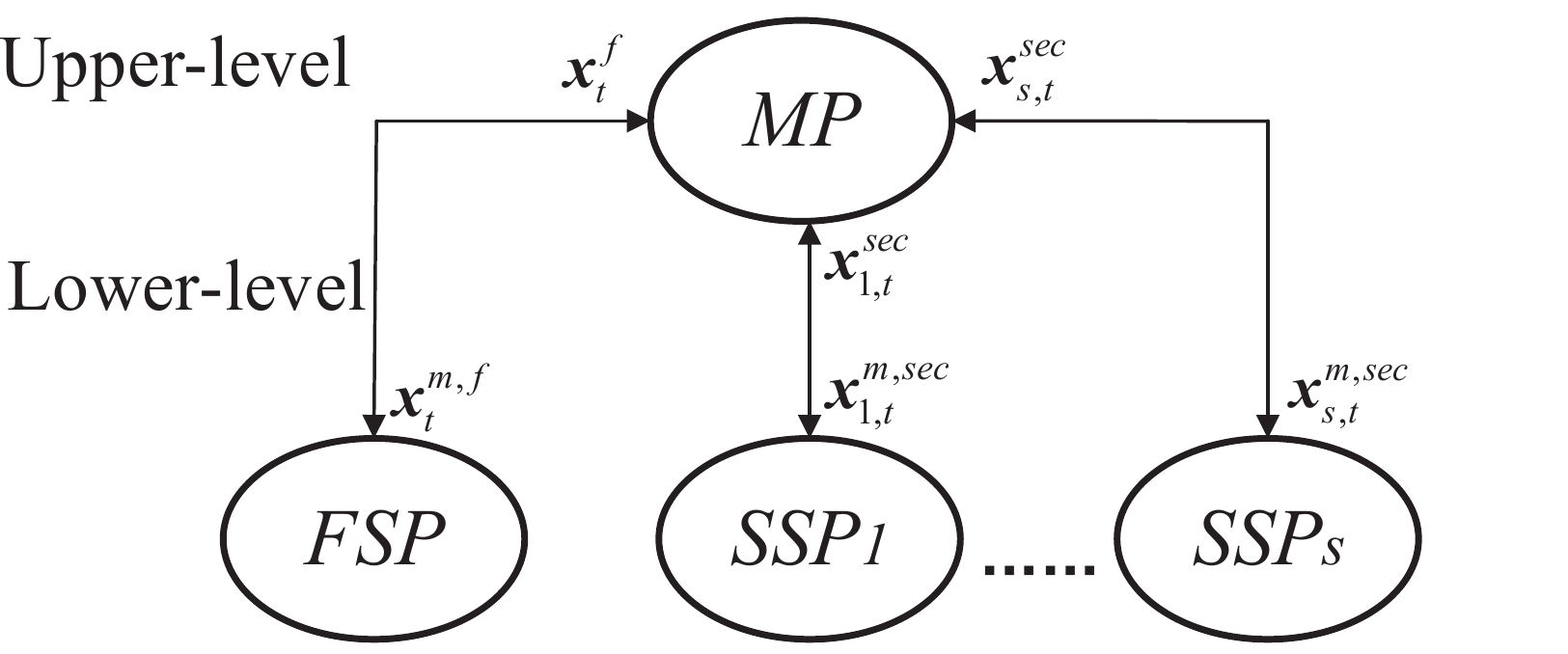}
	\caption{Framework of the ATC-based scenario decomposition algorithm.}
\end{figure}
\section{Scenario Decomposition Algorithm}
\indent{To deal with the above-mentioned drawback, we propose a scenario decomposition method in this section.}\\ 
\subsection{Scenario Decomposition Model}\label{AA}
\indent{Based on the ATC method proposed in \cite{Kargarian2015}, to facilitate the scenario decomposition computation, the original problem can be decomposed into an upper-level master problem and several lower-level subproblems.}\\
\indent{More precisely, the first and the second stage problems are taken as the lower-level subproblems, and a master problem is constructed as the upper-level problem to coordinate the optimization among these two-stage problems. Then, the decision variables of the master problem are regarded as the intermediate variables between the first and second stage variables of problem (1). 
        Hence, the coupling constraint (\ref{eq:eq1d}) can be reformulated into equations (\ref{eq:eq41}) and (\ref{eq:eq42}) as follows:}
   \begin{equation}
      \bm{\tilde{g}}^{f,co}(\bm{x}_{t}^{m,f},\bm{x}_{t}^{f}) \preceq \bm{0}, \forall t \in \mathcal{T} \label{eq:eq41}
   \end{equation}
   \begin{equation}
      \bm{\tilde{g}}^{sec,co}_{s}(\bm{x}_{s,t}^{m,sec},\bm{x}_{s,t}^{sec}) \preceq \bm{0}, \forall t \in \mathcal{T}, \forall s \in \mathcal{S} \label{eq:eq42}
   \end{equation}
\noindent{where $\bm{x}_{t}^{m,f}$ represents the coupling variables (i.e., $P_{DG,i,t}^{f}$ and $Q_{DG,i,t}^{f}$) between the master problem and the first stage problem; $\bm{x}_{s,t}^{m,sec}$ represents the coupling variables (i.e., $P_{DG,s,i,t}$ and $Q_{DG,s,i,t}$) between the master problem and the second stage problem; $ \bm{\tilde{g}}^{f,co}(\cdot)$ represents the coupling function of the master and the first stage variables; $\bm{\tilde{g}}^{sec,co}_{s}(\cdot)$ represents the coupling function of the master and the second stage variables.}\\
\indent{According to \cite{Kargarian2015}, the decomposed models of the master, the first stage, and the second stage problems can be formulated as follows:}\\
\indent{\textit{1) Master Problem (MP)}:}
   \begin{equation}
      \begin{aligned}
      &\min{(\bm{\alpha^{f}})}^{\top}\bm{\tilde{g}}^{f,co}(\bm{x}_{t}^{m,f},\bm{\tilde{x}}_{t}^{f}) \\
      &+ ||\bm{\beta^{f}}\circ\bm{\tilde{g}}^{f,co}(\bm{x}_{t}^{m,f},\bm{\tilde{x}}_{t}^{f})||_{2}^{2}\\
      &+ \sum_{s \in \mathcal{S}}\pi_{s}{(\bm{\alpha^{sec}_{s}})}^{\top}\bm{\tilde{g}}^{sec,co}_{s}(\bm{x}_{s,t}^{m,sec},\bm{\tilde{x}}_{s,t}^{sec})\\
      &+ \sum_{s \in \mathcal{S}}\pi_{s}||\bm{\beta^{sec}_{s}}\circ\bm{\tilde{g}}^{sec,co}_{s}(\bm{x}_{s,t}^{m,sec},\bm{\tilde{x}}_{s,t}^{sec})||_{2}^{2} \label{eq:eq43}
      \end{aligned}
   \end{equation}
\noindent{where $\bm{\alpha^{f}}$ and $\bm{\alpha^{sec}_{s}}$ are Lagrangian multipliers corresponding to equations (\ref{eq:eq41}) and (\ref{eq:eq42}), respectively; $\bm{\beta^{f}}$ and $\bm{\beta^{sec}_{s}}$ are penalty coefficients of augmented Lagrangian functions; $\bm{\tilde{x}}$ represent variables which are parameters in \textit{MP}.}\\
\indent{\textit{2) First Stage Problem (FSP)}:}\\
   \begin{equation}
      \begin{aligned}
      \min C^{f}(\bm{x}_{t}^{f})&+ {(\bm{\alpha^{f}})}^{\top}\bm{\tilde{g}}^{f,co}(\bm{\tilde{x}}_{t}^{m,f},\bm{x}_{t}^{f}) \\
      &+ ||\bm{\beta^{f}}\circ\bm{\tilde{g}}^{f,co}(\bm{\tilde{x}}_{t}^{m,f},\bm{x}_{t}^{f})||_{2}^{2}\\
s.t. & \qquad\qquad\quad(1b)
\label{eq:eq44}
      \end{aligned}
   \end{equation}
\indent{\textit{3) Second Stage Problem (SSP)}:}\\
   \begin{equation}
      \begin{aligned}
      &\min \pi_{s}C^{sec}_{s}(\bm{x}_{s,t}^{sec})\\
      &+ {\pi_{s}(\bm{\alpha^{sec}_{s}})}^{\top}\bm{\tilde{g}}^{sec,co}_{s}(\bm{\tilde{x}}_{s,t}^{m,sec},\bm{x}_{s,t}^{sec})\\
      &+ \pi_{s}||\bm{\beta^{sec}_{s}}\circ\bm{\tilde{g}}^{sec,co}_{s}(\bm{\tilde{x}}_{s,t}^{m,sec},\bm{x}_{s,t}^{sec})||_{2}^{2} \\
&s.t.  \qquad\qquad\qquad\qquad(1c)
\label{eq:eq45}
      \end{aligned}
   \end{equation}
\subsection{ATC-Based Scenario Decomposition Algorithm}
\indent{To find the optimal solutions of the two-stage stochastic model, an ATC-based algorithm is proposed to solve the  aforementioned scenario decomposition models. Note that the global optimality of the algorithm is guaranteed because of the convexity of the studied model \cite{Tosserams2006}. Moreover, the obtained optimal solutions naturally satisfy various kinds of constraints (especially for voltage security limits (9) and (26)) in the first and second stages.}\\
\indent{\textit{1) Framework of the Proposed Algorithm}: The proposed ATC-based scenario decomposition algorithm includes a two-level optimization procedure. Firstly, the lower-level subproblems, i.e., \textit{FSP} and \textit{SSPs}, solve their local problems and send the results, i.e., $\bm{x}_{t}^{f}$ and $\bm{x}_{s,t}^{sec}$, to the upper level. Secondly, the upper-level \textit{MP} solves its local problems and sends decisions, i.e., $\bm{x}_{t}^{m,sec}$ and $\bm{x}_{s,t}^{m,sec}$, to the lower level. Then, the lower-level and upper-level problems update their own parameters and turn to the next iteration. The framework of the proposed method is shown in Fig.1.}\\
\indent{\textit{2) Procedure of the Proposed Algorithm}: The outline of the proposed ATC-based scenario decomposition algorithm is shown in Scheme 1.}\\
\indent{\textit{3) Implementation on Scenarios Parallel Computation}: The aforementioned algorithm is conducted in a serial manner. However, from the model of \textit{FSP} and \textit{SSP}, it can be found that both of them can be solved independently at the same time. Hence, it is possible to conduct the parallel computing in Step 2). The parallel computation implementation framework is described in Fig. 2. In this framework, each thread is assigned to solve a subproblem (i.e., \textit{MP}, \textit{FSP} or \textit{SSP}). Furthermore, all the threads corresponding to \textit{FSP} and \textit{SSPs} are running in parallel during each iteration. Note that parallel computing makes more sense when the number of scenarios becomes large. }\\
\vspace{-0.4cm}
\begin{flushleft}
   \begin{tabular*}{25em}%
   {@{\extracolsep{\fill}}l }
   \hline
   \bf{Scheme 1: ATC-based scenario decomposition algorithm}\\
   \hline
   \noindent{\bf{1: Initialization:}}\noindent{ Set the iteration index $a=1$, the initial pa-}\\ 
   \hspace{0.3cm}\noindent{ rameters $\bm{\alpha^{f,a},\bm{\beta^{f,a}}, \bm{\alpha^{sec,a}_{s}}}, \bm{\beta^{sec,a}_{s}},\lambda$, and the conver- }\\
   \hspace{0.45cm}{gence tolerance $\varepsilon $.}\\
   \noindent{\bf{2: Lower-level optimization:}}\\
   \hspace{0.45cm}\textbf{for}\noindent{\textit{ FSP}:}\\
   \hspace{0.6cm}\noindent{2.1: Solve \textit{FSP}};\\
   \hspace{0.6cm}\noindent{2.2: Send $\bm{x}_{t}^{f,a}$ to \textit{MP}};\\
   \hspace{0.45cm}\textbf{for}\noindent{ all \textit{SSPs}:}\\
   \hspace{0.6cm}\noindent{2.3: Solve \textit{SSP};}\\
   \hspace{0.6cm}\noindent{2.4: Send $\bm{x}_{s,t}^{sec,a}$ to \textit{MP}};\\
   \noindent{\bf{3: Upper-level optimization:}}\\
   \hspace{0.45cm}\textbf{for}\noindent{ \textit{MP}:}\\
   \hspace{0.6cm}\noindent{3.1: Solve \textit{MP}};\\
   \hspace{0.6cm}\noindent{3.2: Send $\bm{x}_{t}^{m,f,a}, \bm{x}_{s,t}^{m,sec,a}$ to \textit{FSP} and \textit{SSPs}, respectively};\\
   \noindent{\bf{4: Parameter update:}}\\
   \hspace{0.45cm}\textbf{for}\noindent{\textit{ FSP}:}\\
   \hspace{0.6cm}\noindent{4.1: Receive $\bm{x}_{t}^{m,f,a}$ from \textit{MP}};\\
   \hspace{0.6cm}\noindent{4.2: Update and fix $\bm{\tilde{x}}_{t}^{m,f,a}$ in the model};\\
   \hspace{0.45cm}\textbf{for}\noindent{ all\textit{ SSPs}:}\\
   \hspace{0.6cm}\noindent{4.3: Receive $\bm{x}_{s,t}^{m,sec,a}$ from \textit{MP}};\\
   \hspace{0.6cm}\noindent{4.4: Update and fix $\bm{\tilde{x}}_{s,t}^{m,sec,a}$ in the model};\\
   \hspace{0.45cm}\textbf{for}\noindent{\textit{ MP}:}\\
   \hspace{0.6cm}\noindent{4.5: Receive $\bm{x}_{t}^{f,a},\bm{x}_{s,t}^{sec,a}$ from \textit{FSP} and \textit{SSPs}, respectively};\\
   \hspace{0.6cm}\noindent{4.6: Update and fix $\bm{\tilde{x}}_{t}^{f,a},\bm{\tilde{x}}_{s,t}^{sec,a}$ in the model};\\
   \hspace{0.6cm}\noindent{4.7: Update parameters according to (46)-(49):}\\
   \hspace{0.9cm}$\bm{\alpha^{f,a+1}}=\bm{\alpha^{f,a}} + 2(\bm{\beta^{f,a}})^{2}(\bm{x}_{t}^{m,f,a}-\bm{\tilde{x}}_{t}^{f,a})\quad$\hspace{0.7cm}\noindent{(46)}\\
   \hspace{0.9cm}$\bm{\beta^{f,a+1}}=\lambda\bm{\beta^{f,a}}$\noindent{~~~~~~~~~~~~~~~~~~~~~~~~~~~~~~~~~~~~~~~~(47)}\\
   \hspace{0.6cm}$\bm{\alpha^{sec,a+1}_{s}}=\bm{\alpha^{sec,a}_{s}} + 2(\bm{\beta^{sec,a}_{s}})^{2}(\bm{x}_{s,t}^{m,sec,a}-\bm{\tilde{x}}_{s,t}^{sec,a})$\noindent{(48)}\\
   \hspace{0.6cm}$\bm{\beta^{sec,a+1}_{s}}=\lambda\bm{\beta^{sec,a}_{s}}$\noindent{~~~~~~~~~~~~~~~~~~~~~~~~~~~~~~~~~~~~~~(49)}\\
   \hspace{0.6cm}\noindent{4.8: Send $(\bm{\alpha^{f,a+1}}, \bm{\beta^{f,a+1}})$ and $(\bm{\alpha^{sec,a+1}_{s}}, \bm{\beta^{sec,a+1}_{s}})$}\\
   \hspace{1.25cm}\noindent{ to \textit{FSP} and \textit{SSPs}, respectively}\\
   \noindent{\bf{5: Convergence check: }}\\
   \hspace{0.6cm}\noindent{5.1: If (50) is satisfied, stop the algorithm;}\\
   \hspace{1.1cm}$max(|\bm{x}_{t}^{m,f,a}-\bm{\tilde{x}}_{t}^{f,a}|,|\bm{x}_{s,t}^{m,sec,a}-\bm{\tilde{x}}_{s,t}^{sec,a}|) \le \varepsilon$\hspace{0.5cm}\noindent{(50)}\\
   \hspace{0.6cm}\noindent{5.2: Otherwise,  set $a \leftarrow a+1$ and return to Step 2.}\\
   \hline
   \end{tabular*}
\end{flushleft}
\begin{figure}[!]
   \centering
   \setlength{\abovecaptionskip}{0cm}
   \includegraphics[scale=0.8]{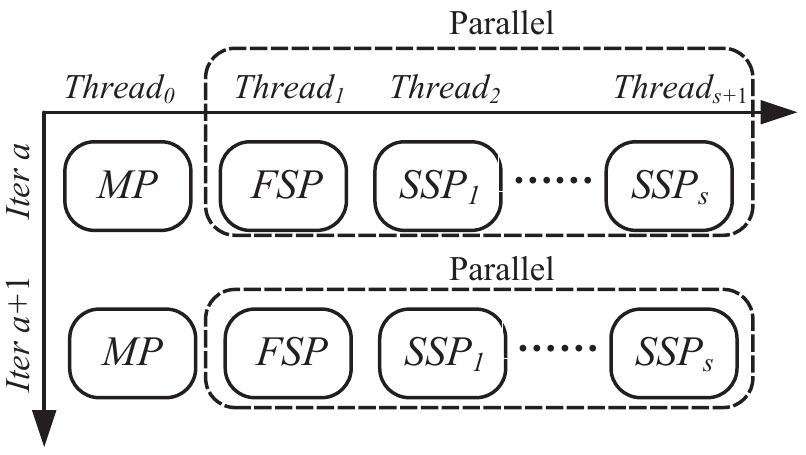}
   \caption{Scenario parallel computation implementation framework.}
\end{figure}
\section{Case Study}
\indent{In order to test the effectiveness of the proposed algorithm, several case studies are carried out on the IEEE 33-bus distribution network and various larger systems. These distribution networks are assumed to operate independently from the upper-level grid. }\\
\indent{All of the optimization models are solved by using GUROBI on a personal computer with an Intel Core i5 CPU and 8GB memory. The whole optimization period is set to 24 hours with dispatch interval 1 hour. To test the accuracy of the proposed algorithm, scenario-centralized method is used as the benchmark. This means that the original problem (1) is solved directly without decomposition.}
	\begin{figure}[!]
	   \centering
	   \setlength{\abovecaptionskip}{0cm}
	   \includegraphics[scale=0.35]{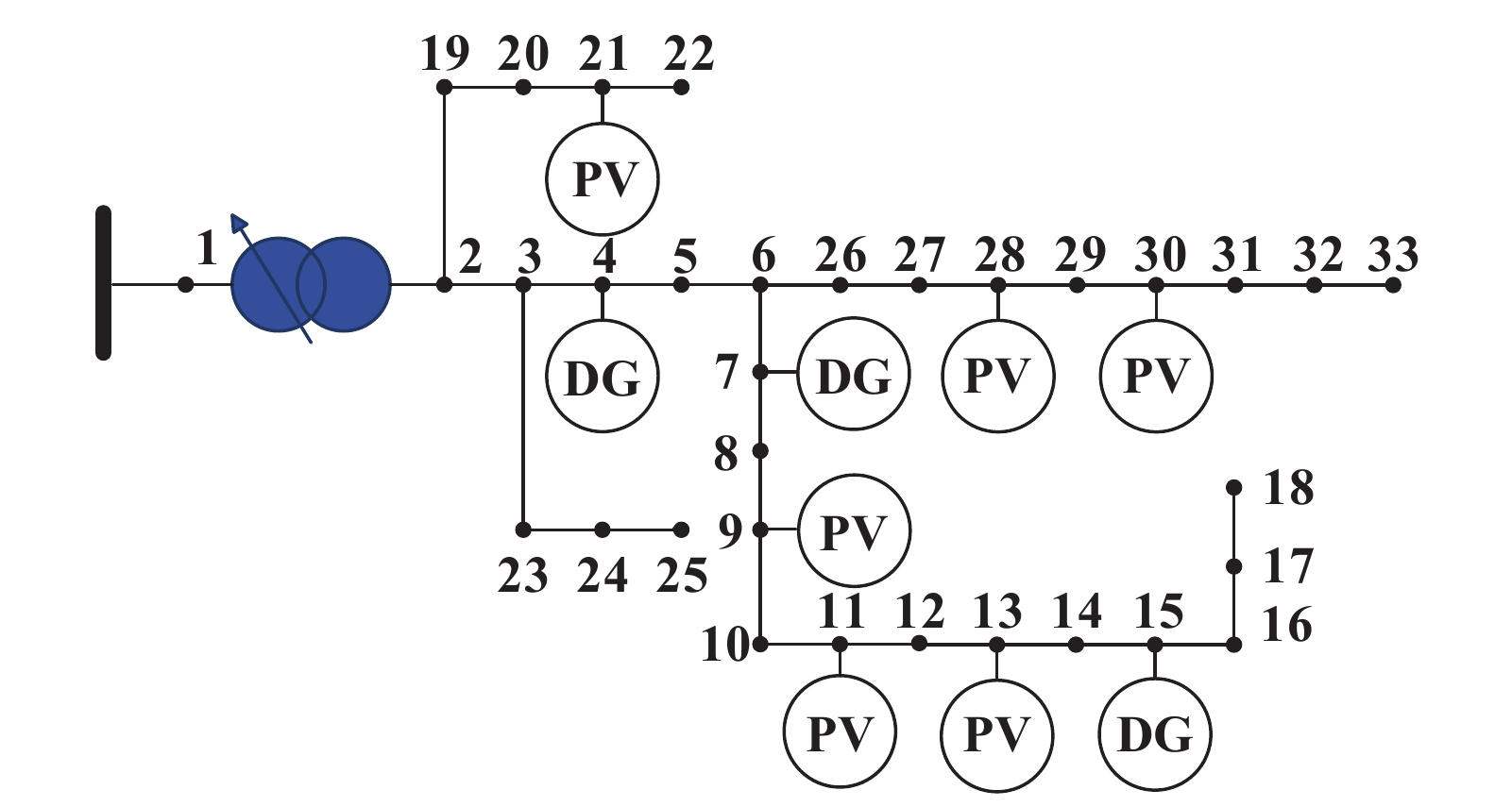}
	   \caption{Configuration of the IEEE 33-bus system.}
	   \vspace{-0.4cm}
	\end{figure}
   \begin{figure}[!]
      \setlength{\abovecaptionskip}{0cm}
      \includegraphics[scale=0.28]{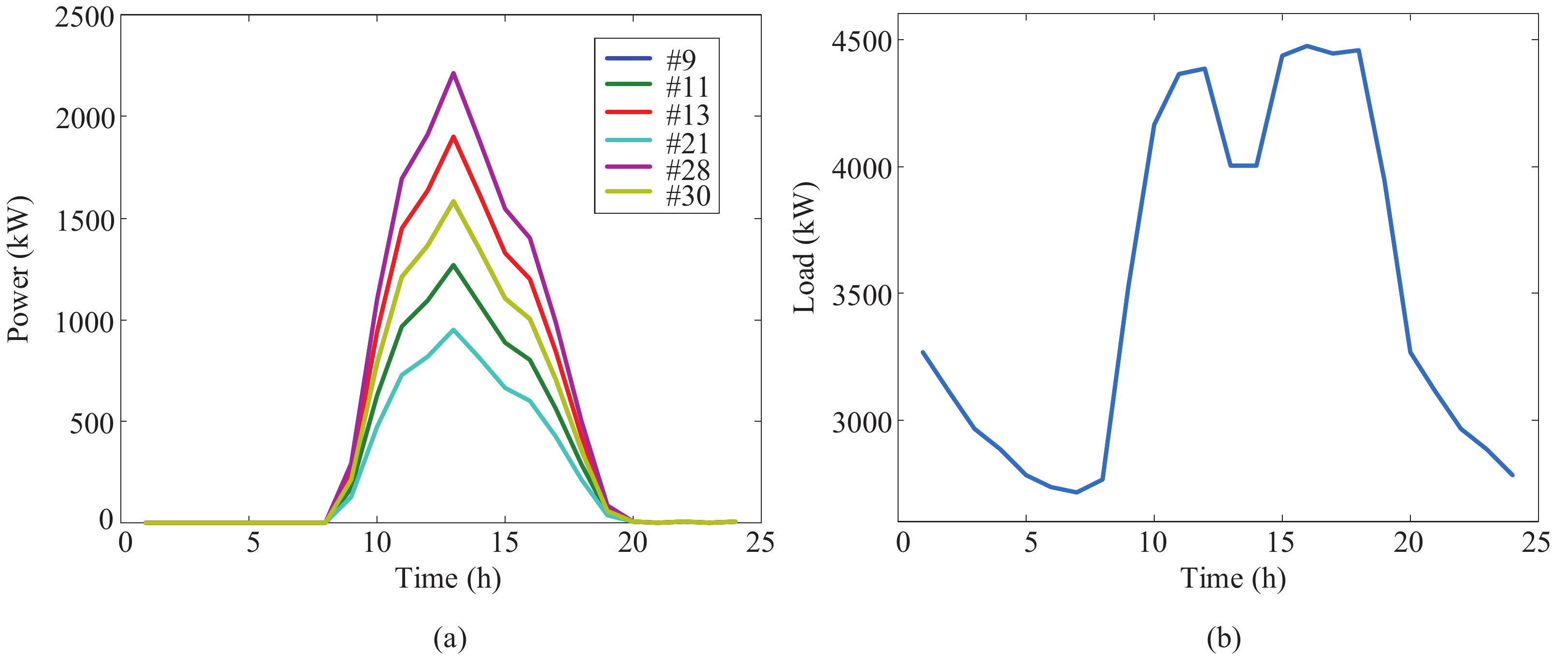}
      \caption{Configuration of system parameters. (a) Forecasted scenario curve of PVs. (b) Forecasted scenario curve of loads}
      \vspace{-0.4cm}
   \end{figure}
\subsection{Deterministic Case}
\indent{Firstly, the proposed algorithm is tested on the IEEE 33-bus system under the deterministic case. The system configuration is shown in Fig. 3. The capacities of DGs at Bus 4, 7, and 15 are set to [0.14, 1.4], [0.12, 1.2], and [0.15, 1.5] MW, respectively. The capacities of PVs at Bus 9, 11, 13, 21, 28, 30 are set from 0 to 1.6, 1.4, 2, 1, 2.5, and 1.6 MW, respectively. The forecasted scenarios of PVs and loads are shown in Fig. 4. We assume that PVs adopt maximum power point tracking technique. Considering lower limits of DGs, the maximum PV outputs can be up to 92.20\% of the total load. The power factors are set to 0.9 for all load buses.

It is assumed that there is only one error scenario with probability one in the second-stage problem. Hence, the influences of stochastic factors (i.e., the uncertain available PV outputs and loads) are ignored. The algorithm parameter settings are $\alpha=\beta=2$. Convergence tolerance $\varepsilon$ is set to 0.0005.}\\
\indent{The simulation results are shown in Table I, Fig. 5 and Fig. 6. It can be found that the optimal solutions (i.e., costs, power losses and dispatched power outputs of DGs and PVs per hour) of the benchmark and the proposed algorithm are almost the same. This also shows that the optimal dispatch results can be made in a decomposed manner. }\\
\indent{Note that the computation time of the proposed algorithm is longer than the benchmark. The reason is that the proposed algorithm does not focus on better computation efficiency compared with the benchmark but on solving complex cases with a large number of scenarios. 
        Moreover, it is also an interesting topic on achieving the optimal solutions in a shorter time than the benchmark method.}\\
   \begin{table}[!]
      \renewcommand{\arraystretch}{1.3}
      \caption{Simulation Results of the Deterministic Case}
      \label{Tab1}
      \centering
      \vspace{-0.3cm}
      \begin{tabular}{c|c|c}
      \hline
       & Benchmark & Proposed algorithm  \\
      \hline
      First-stage Costs (\$) & 247.21 & 247.22 \\
      \hline
      Second-stage Costs (\$) & 354.47 & 354.47 \\
      \hline
      Total Costs (\$) & 601.68 & 601.69 \\
      \hline
      Iterations & $-$ & 96 \\
      \hline
      Time (s) & 0.65 & 84.30 \\
      \hline
      Obj-Err (\%) & $-$ & 0.00 \\
      \hline
      \end{tabular}
      \vspace{-0.4cm}
   \end{table}
	\begin{figure}[!]
	   \centering
	   \setlength{\abovecaptionskip}{0cm}
	   \includegraphics[scale=0.28]{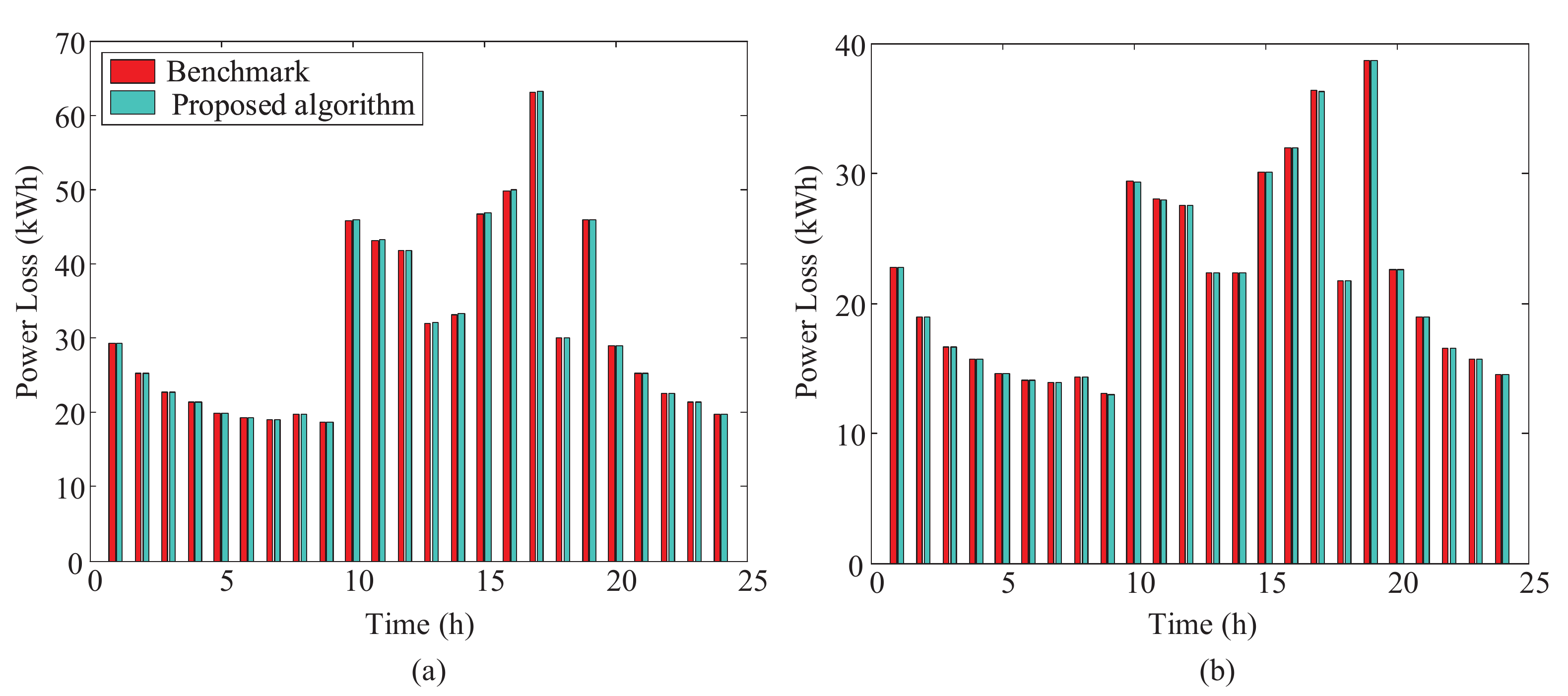}
	   \caption{Comparison of optimization solutions. (a) The first-stage power losses. (b) The second-stage power losses.}
	   \vspace{-0.5cm}
	\end{figure}
   \begin{figure}[!]
      \setlength{\abovecaptionskip}{0cm}
      \includegraphics[scale=0.28]{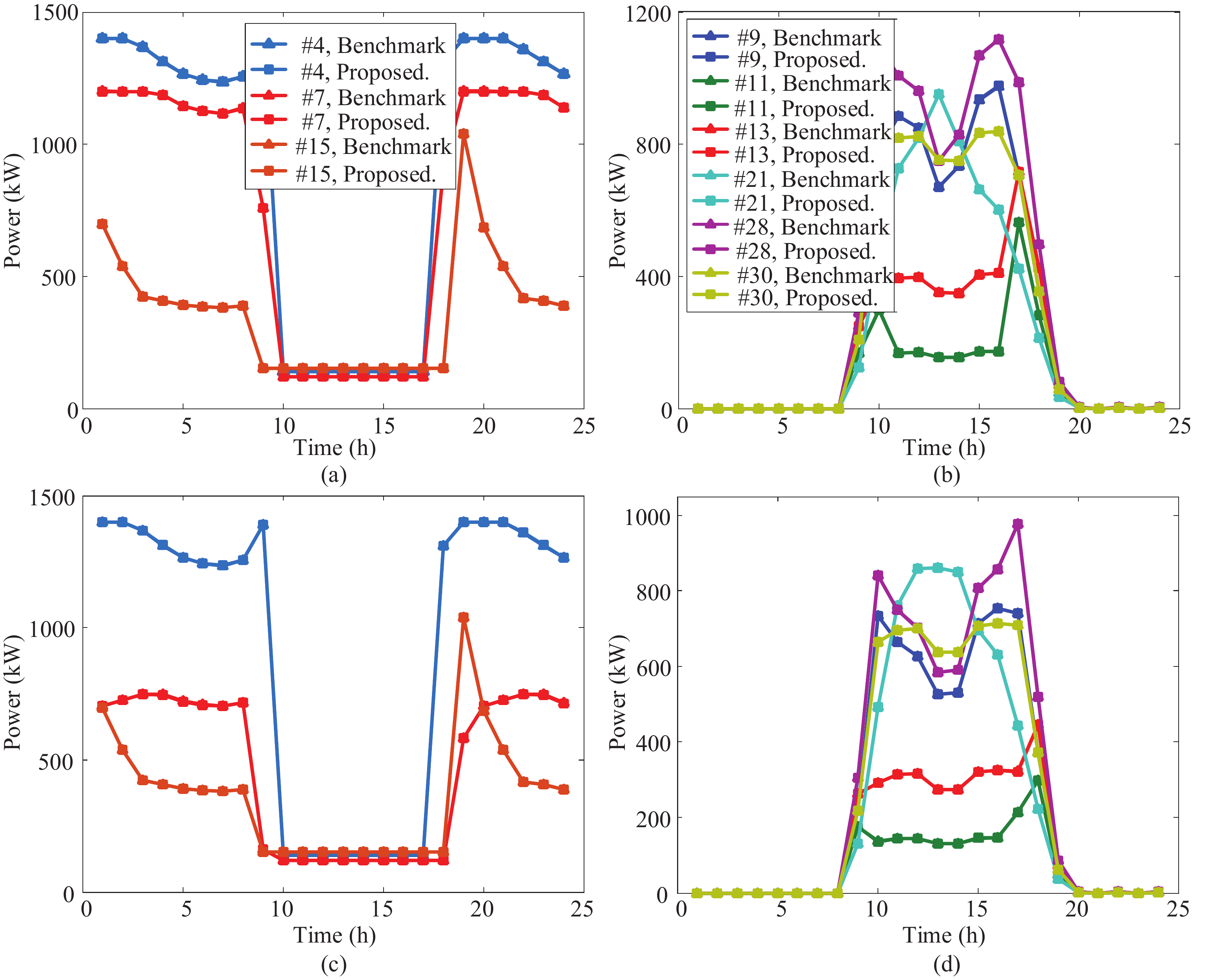}
      \caption{Comparison of the power outputs between the benchmark and the proposed algorithm. 
      (a) First-stage DGs' outputs. (b) First-stage PVs' outputs. (c) Second-stage DGs' outputs. (d) Second-stage PVs' outputs.}
      \vspace{-0.5cm}
   \end{figure}
   \begin{figure}[!]
      \centering
      \setlength{\abovecaptionskip}{0cm}
      \includegraphics[scale=0.41]{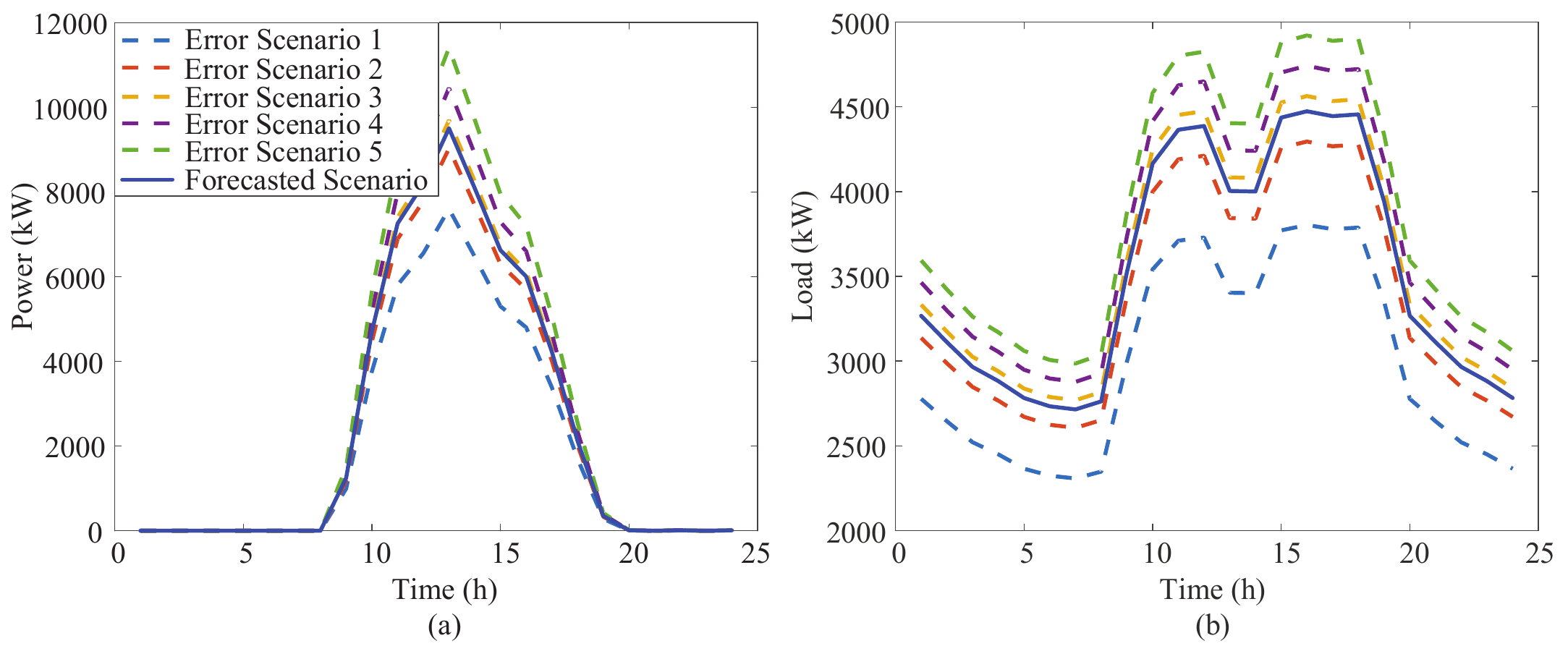}
      \caption{Illustration of scenario profiles for $|\mathcal{S}|=5$ case. (a) Scenario profiles of PVs. (b) Scenario profiles of loads.}
      \vspace{-0.5cm}
   \end{figure}
   \begin{table}[!]
      \renewcommand{\arraystretch}{1.3}
      \caption{Simulation Settings for Error Scenarios}
      \label{Tab2}
      \centering
      \vspace{-0.3cm}
      \begin{tabular}{c|c|c|c|c|c}
      \hline
      \multicolumn{6}{c}{$|\mathcal{S}|=5$}\\
      \hline
       & $s1$ & $s2$ & $s3$ & $s4$ & $s5$  \\
      \hline
       $\pi_{s}$ & 0.1 & 0.2 & 0.4 & 0.2 & 0.1  \\
      \hline
      $\rho_{load}$ & 0.85 & 0.96 & 1.02 & 1.06 & 1.1 \\
      \hline
      $\rho_{pv}$ & 0.8 & 0.95 & 1.02& 1.1 & 1.2  \\
      \hline
      \multicolumn{6}{c}{$|\mathcal{S}|=10$}\\
      \hline
       & $s1$ & $s2$ & $s3$ & $s4$ & $s5$  \\
      \hline
      $\pi_{s}$ & 0.02 & 0.03 & 0.05 & 0.1 & 0.3  \\
      \hline
      $\rho_{load}$ & 0.85 & 0.9 & 0.95 & 0.96 & 0.98 \\
      \hline
      $\rho_{pv}$ & 0.8 & 0.85 & 0.9& 0.95 & 0.98  \\
      \hline
      & $s6$ & $s7$ & $s8$ & $s9$ & $s10$  \\
      \hline
      $\pi_{s}$ & 0.3 & 0.1 & 0.05 & 0.03 & 0.02  \\
      \hline
      $\rho_{load}$ & 1.02 & 1.04 & 1.06 & 1.07 & 1.1 \\
      \hline
      $\rho_{pv}$ & 1.02 & 1.05 & 1.1& 1.15 & 1.2  \\
      \hline
      \multicolumn{6}{c}{$|\mathcal{S}|=20$}\\
      \hline
       & $s1$ & $s2$ & $s3$ & $s4$ & $s5$  \\
      \hline
      $\pi_{s}$ & 0.001 & 0.002 & 0.0045 & 0.01 & 0.02  \\
      \hline
      $\rho_{load}$ & 0.85 & 0.86 & 0.87 & 0.88 & 0.9 \\
      \hline
      $\rho_{pv}$ & 0.8 & 0.81 & 0.82& 0.83 & 0.85  \\
      \hline
      & $s6$ & $s7$ & $s8$ & $s9$ & $s10$  \\
      \hline
      $\pi_{s}$ & 0.05 & 0.1 & 0.15 & 0.3 & 0.15  \\
      \hline
      $\rho_{load}$ & 0.92 & 0.95 & 0.96 & 0.97 & 0.98 \\
      \hline
      $\rho_{pv}$ & 0.86 & 0.87 & 0.9& 0.92 & 0.95  \\
      \hline
      & $s11$ & $s12$ & $s13$ & $s14$ & $s15$  \\
      \hline
      $\pi_{s}$ & 0.1 & 0.05 & 0.02 & 0.015 & 0.01  \\
      \hline
      $\rho_{load}$ & 0.99 & 1.02 & 1.03 & 1.04 & 1.05 \\
      \hline
      $\rho_{pv}$ & 0.97 & 0.98 & 1.01& 1.02 & 1.05  \\
      \hline
      & $s16$ & $s17$ & $s18$ & $s19$ & $s20$  \\
      \hline
      $\pi_{s}$ & 0.008 & 0.004 & 0.003 & 0.0015 & 0.001  \\
      \hline
      $\rho_{load}$ & 1.06 & 1.07 & 1.09 & 1.1 & 1.12 \\
      \hline
      $\rho_{pv}$ & 1.07 & 1.1 &1.12& 1.15 & 1.2  \\
      \hline
      \end{tabular}
      \begin{tablenotes}
         \item[] Notation: Denote $\rho_{load}$ and $\rho_{pv}$ as error scenario factors of loads and PVs, respectively. Denote $s1$ as scenario 1.
      \end{tablenotes}
   \end{table}
   \begin{figure}[!]
      \centering
      \setlength{\abovecaptionskip}{0cm}
      \includegraphics[scale=0.32]{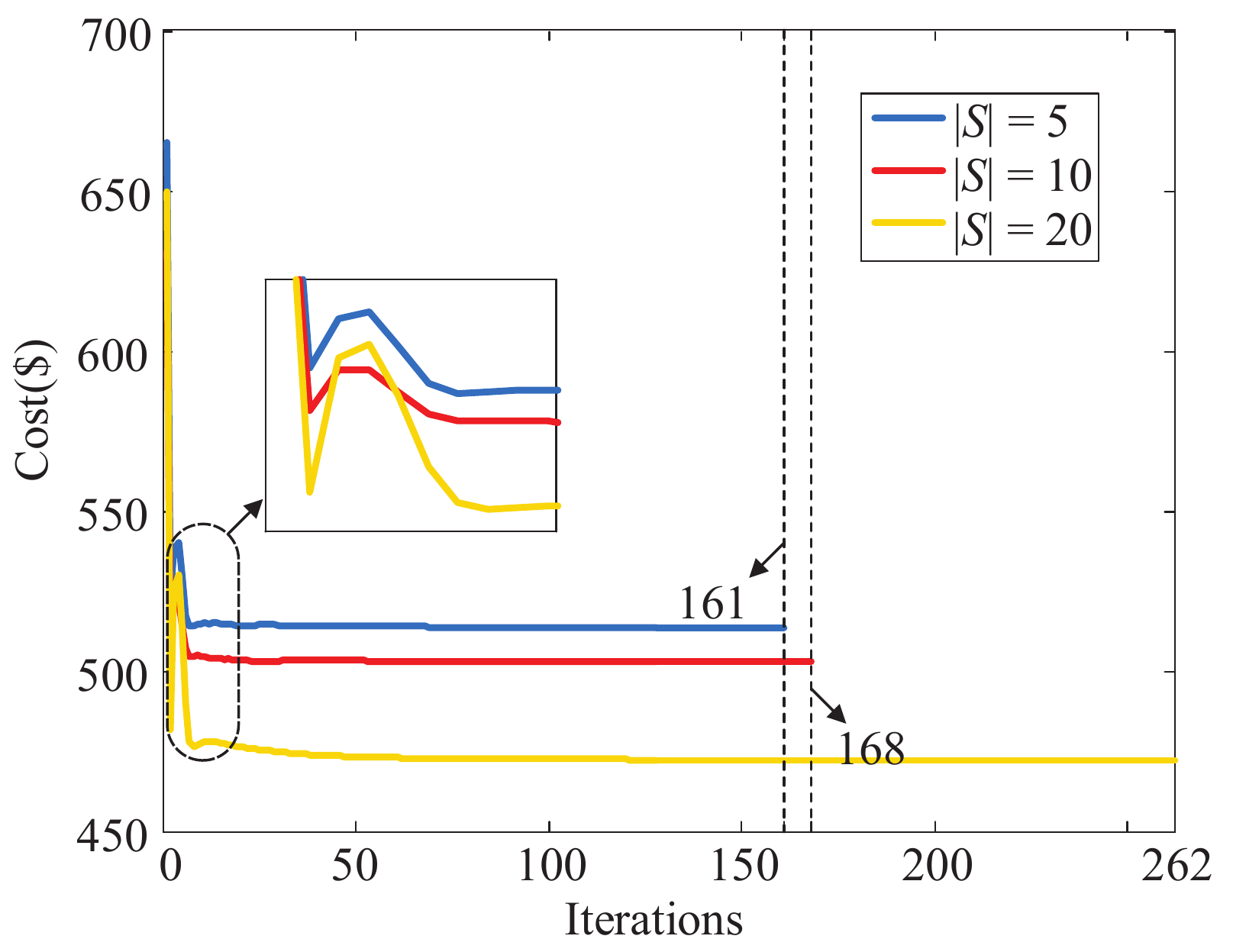}
      \caption{Convergence procedure of the proposed algorithm.}
      \vspace{-0.6cm}
   \end{figure}
\vspace{-0.5cm}
\subsection{Stochastic Case}
\indent{In this section, the effectiveness and adaptability of the proposed algorithm are further tested on the IEEE 33-bus system and several larger systems. We consider the stochastic characteristics of available PV outputs and loads, then test the proposed algorithm under multiple error scenarios (i.e., scenario number $|\mathcal{S}|=5$, $|\mathcal{S}|=10$, and $|\mathcal{S}|=20$) with different probabilities (the sum equals one). Each error scenario can be obtained by multiplying the forecasted scenario shown in Fig. 4 with the corresponding factor $\rho$ in Table II. Detailed simulation settings for error scenarios are summarized in Table II. For brevity, we only depict scenario profiles of $|\mathcal{S}|=5$ which are shown in Fig. 7. For other scenario cases, the scenario profiles can be obtained in the similar way.}\\
\indent{\textit{1) Effectiveness on the IEEE 33-Bus System}: To better illustrate the effectiveness of the proposed algorithm, we test the system with different numbers of scenarios, i.e., $|\mathcal{S}|=5,10$, and $20$. The parameter settings are $\alpha=\beta=3.5$ for $|\mathcal{S}|=5$, $\alpha=\beta=3$ for $|\mathcal{S}|=10$, and $\alpha=\beta=4.5$ for $|\mathcal{S}|=20$. Convergence tolerance $\varepsilon$ is set to 0.0005. Moreover, the lower-level parallel computing version of the proposed algorithm is also implemented to demonstrate the application potential in the case of a larger number of scenarios.}\\
\indent{From Table III and Fig. 8, it can be concluded that, as the number of scenarios increases, the speed-up effect of parallel computing becomes more obvious. Moreover, from Table III and Fig. 9, the accuracy of the proposed algorithm compared with the benchmark is also preserved with the different number of scenarios. This further demonstrates the effectiveness of the proposed algorithm on dealing with stochastic optimization in a decomposed manner. }\\
\indent{\textit{2) Adaptability in Larger Systems}: To demonstrate the adaptability of the proposed method, we further conduct simulations on various large systems by duplicating the IEEE 33-bus system. These systems include the 330-bus system with scenario number $|\mathcal{S}|=5$, the 660-bus system with scenario number $|\mathcal{S}|=10$, the 1320-bus system with scenario number $|\mathcal{S}|=20$, and the 1980-bus system with scenario number $|\mathcal{S}|=20$. Convergence tolerance $\varepsilon$ is set to 0.02. And all the other parameter settings are identical to the case in Section IV.B.\textit{1)}.}\\
\indent{From Table IV, it can be found that, with the increase in system scale and scenario numbers, the benchmark method failed to obtain the solution for the 1320 system with $|\mathcal{S}|=20$ and the 1980 system with $|\mathcal{S}|=20$. The reason is that the scale of the scenario-centralized model becomes too large and exceeds the upper limit of computer memory. However, the proposed ATC-based scenario-decomposition method is still suitable for solving such large-scale cases. This further demonstrates the good adaptability of the proposed method. }
   \begin{table}[!]
      \renewcommand{\arraystretch}{1.3}
      \caption{Simulation Results of the Stochastic Case on the IEEE 33-Bus System}
      \label{Tab2}
      \centering
      \vspace{-0.3cm}
      \begin{tabular}{c|c|c|c|c}
      \hline
       &   & Benchmark & Serial & Parallel \\
      \hline
       \multirow{5}*{$|\mathcal{S}|=5$}   & Costs(\$) & 513.57 & 513.61 & 513.61  \\
      \cline{2-5}
                & Obj-Err (\%) & $-$ & 0.01 & 0.01  \\
      \cline{2-5}
                & Iterations & $-$ & 161 & 161  \\
      \cline{2-5}
                & Time (s) & 1.35 & 405.24 & 156.84 \\
      \cline{2-5}
                & Speed-up (\%) & $-$ & $-$ & 61.30  \\
      \hline
      \multirow{5}*{$|\mathcal{S}|=10$}   & Costs(\$) & 502.59 & 502.99 & 502.99 \\
      \cline{2-5}
                & Obj-Err (\%) & $-$ & 0.08 & 0.08  \\
      \cline{2-5}
                & Iterations & $-$ & 168 & 168  \\
      \cline{2-5}
                & Time (s) & 3.06 & 850.86 & 277.19 \\
      \cline{2-5}
                & Speed-up (\%) & $-$ & $-$ & 67.42  \\
      \hline
      \multirow{5}*{$|\mathcal{S}|=20$}   & Costs(\$) & 472.04 & 472.17 & 472.17 \\
      \cline{2-5}
                & Obj-Err (\%) & $-$ & 0.03 & 0.03  \\
      \cline{2-5}
                & Iterations & $-$ & 262 & 262  \\
      \cline{2-5}
                & Time (s) & 6.73 & 2889.59 & 780.19 \\
      \cline{2-5}
                & Speed-up (\%) & $-$ & $-$ & 72.69  \\
      \hline
      \end{tabular}
      \begin{tablenotes}
         \item[] Notation: Denote the serial and lower-level parallel versions as Serial and Parallel, respectively.
      \end{tablenotes}
      \vspace{-0.4cm}
   \end{table}
\begin{figure}[!]
   \centering
   \setlength{\abovecaptionskip}{0cm}
   \includegraphics[scale=0.42]{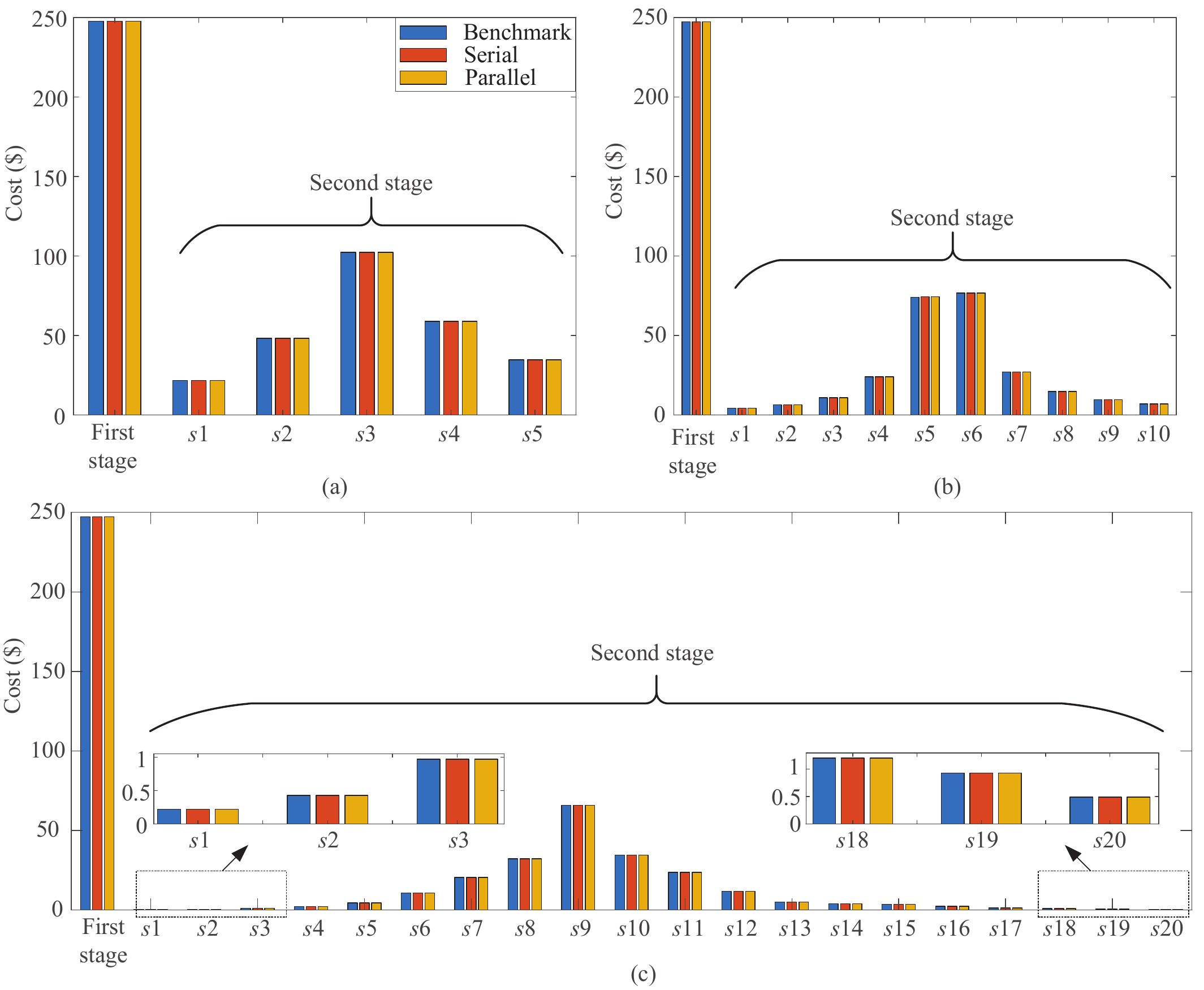}
   \caption{Optimization costs of the stochastic case on the IEEE 33-bus system. (a) Costs of $|\mathcal{S}|=5$ case. (b) Costs of $|\mathcal{S}|=10$ case. (c) Costs of $|\mathcal{S}|=20$ case.\\
   Notation: Denote $s$1 as scenario 1.}
   \vspace{-0.6cm}
\end{figure}
\begin{table}[!]
   \renewcommand{\arraystretch}{1.3}
   \caption{Simulation Results of the Stochastic Case on Larger Systems}
   \label{Tab2}
   \centering
   \vspace{-0.3cm}
   \begin{tabular}{m{1.5cm}<{\centering}|c|c|c|c}
   \hline
    &   & Benchmark & Serial & Parallel \\
   \hline
    \multirow{5}*{\makecell{330-Bus\\$|\mathcal{S}|=5$}}   & Costs(\$) & 5100.94 & 5102.38 & 5102.38\\
   \cline{2-5}
             & Obj-Err (\%) & $-$ & 0.03 & 0.03  \\
   \cline{2-5}
             & Iterations & $-$ & 130 & 130  \\
   \cline{2-5}
             & Time (s) & 18.29 & 2934.31 & 1447.342 \\
   \cline{2-5}
             & Speed-up (\%) & $-$ & $-$ & 50.68  \\
   \hline
   \multirow{5}*{\makecell{660-Bus\\$|\mathcal{S}|=10$}}   & Costs(\$) & 9958.64 & 9967.34 & 9967.34 \\
   \cline{2-5}
             & Obj-Err (\%) & $-$ & 0.09 & 0.09  \\
   \cline{2-5}
             & Iterations & $-$ & 156 & 156  \\
   \cline{2-5}
             & Time (s) & 89.25 & 15066.17 & 7137.56 \\
   \cline{2-5}
             & Speed-up (\%) & $-$ & $-$ & 52.63  \\
   \hline
   \multirow{5}*{\makecell{1320-Bus\\$|\mathcal{S}|=20$}}   & Costs(\$) & Unsolvable & 18678.56 & 18678.56 \\
   \cline{2-5}
             & Obj-Err (\%) & $-$ & $-$ & $-$  \\
   \cline{2-5}
             & Iterations & $-$ & 280 & 280  \\
   \cline{2-5}
             & Time (s) & $-$ & 128823.43 & 58148.63 \\
   \cline{2-5}
             & Speed-up (\%) & $-$ & $-$ & 54.86  \\
   \hline
   \multirow{5}*{\makecell{1980-Bus\\$|\mathcal{S}|=20$}}   & Costs(\$) & Unsolvable & 28004.14 & 28004.14 \\
   \cline{2-5}
             & Obj-Err (\%) & $-$ & $-$ & $-$  \\
   \cline{2-5}
             & Iterations & $-$ & 278 & 278  \\
   \cline{2-5}
             & Time (s) & $-$ & 212014.48 & 94112.76 \\
   \cline{2-5}
             & Speed-up (\%) & $-$ & $-$ & 55.61  \\
   \hline
   \end{tabular}
   \begin{tablenotes}
      \item[] Notation: Denote the serial and lower-level parallel versions as Serial and Parallel, respectively.
   \end{tablenotes}
   \vspace{-0.7cm}
\end{table}

\section{Conclusion}
\indent{In this paper, the ATC-based scenario decomposition algorithm has been proposed to deal with the two-stage stochastic OPF model. Some conclusions have been obtained as follows:}\\
\indent{1) The proposed algorithm significantly reduces the complexity of the traditional scenario-based model and effectively overcome the curse of dimensionality. For complex cases that cannot be solved by the traditional method, the proposed method is still applicable.}\\
\indent{2) The proposed decomposition algorithm remains high accuracy, whose solutions are very close to the benchmark method.}\\
\indent{3) The parallel computing can be implemented among the lower-level subproblems corresponding to scenarios. With the increase of scenario numbers, the acceleration effect of parallel computing becomes more obvious. It shows that the proposed algorithm has good potential for applications.}
\vspace{-0.1cm}




%

\end{document}